\documentclass[aps,pre,twocolumn,superscriptaddress,showpacs]{revtex4}
\usepackage{amsmath,epsfig,amsfonts,amssymb,latexsym,times}
\begin{document}
\title{Randomness enhances cooperation: \\a resonance type phenomenon in evolutionary games}

\author{Jie Ren$^{1}$}
\author{Wen-Xu Wang$^{2}$}
\author{Feng Qi$^{3}$}\email{fqi@mcw.edu}
\affiliation{$^{1}$Department of Physics, University of Fribourg,
CH-1700, Fribourg, Switzerland\\
$^{2}$Department of Electronic Engineering, City University of Hong Kong, Hong Kong SAR, China\\
$^{3}$Biotechnology and Bioengineering Center and Department of
Physiology, Medical College of Wisconsin, Milwaukee, Wisconsin
53226, USA}
\date{\today}

\begin{abstract}
We investigate the effect of randomness in both relationships and
decisions on the evolution of cooperation. Simulation results
show, in such randomness' presence, the system evolves more
frequently to a cooperative state than in its absence.
Specifically, there is an optimal amount of randomness, which can
induce the highest level of cooperation. The mechanism of
randomness promoting cooperation resembles a resonance-like
fashion, which could be of particular interest in evolutionary
game dynamics in economic, biological and social systems.
\end{abstract}
\maketitle

Cooperation is ubiquitous in the real world, ranging from
biological systems to economic and social systems. However, the
unselfish, altruistic actions apparently contradict Darwinian
selection. Thus, understanding the conditions for the emergence
and maintenance of cooperative behavior among selfish individuals
becomes a central issue \cite{cooperate1}. In the last decades,
several natural mechanisms of enforcing cooperation have already
been explored such as kin selection \cite{kin}, retaliating
behavior \cite{Axelrod}, reciprocity \cite{Nowak1}, voluntary
participation \cite{Hauert1}, development of reputation
\cite{reputation}, or spatial extensions \cite{Nowak}.

Since the pioneering work on iterated games by Axelrod
\cite{Axelrod}, the evolutionary Prisoner's Dilemma Game (PDG) as
a general metaphor for studying cooperative behavior has drawn
much attention from scientific communities. Szab\'{o} presented a
stochastic evolutionary rule to capture the bounded rationality of
individuals for better characterizing the dynamics of games in
real systems \cite{Szabo5}. The individuals can follow only two
simple strategies: $C$ (cooperate) or $D$ (defect), described by
\begin{equation}
s=\left(
  \begin{array}{c}
    1 \\
    0 \\
  \end{array}
\right) or \left(
  \begin{array}{c}
    0 \\
    1 \\
  \end{array}
\right)
\end{equation}
respectively. Each individual plays the PDG with its ``neighbors"
defined by their spatial relationships. The total income of the
player at the site $x$ can be expressed as
\begin{equation}
M_x=\sum_{y\in\Omega_x}s_x^{T}\cdot P\cdot s_y
\end{equation}
where $s_x$ and $s_y$ denote the strategy of node $x$ and $y$. The
sum runs over all the neighboring sites of $x$ (this set is
indicated by $\Omega_x$) and the payoff matrix has a rescaled form
suggested by Nowak and May \cite{May}:
\begin{equation}
P=\left(
    \begin{array}{cc}
      1 & 0 \\
      b & 0 \\
    \end{array}
  \right)
\end{equation}
where $1<b<2$. Then, the individual $x$ randomly selects a
neighbor $y$ for possible updating its strategy. The probability
that $x$ follows the strategy of the selected node $y$ is
determined by the total payoff difference between them:
\begin{equation}
\label{eq:tpo}
W_{s_x\leftarrow s_y}=\frac{1}{1+\exp[(M_x-M_y)/T]},
\end{equation}
where $T$ characterizes the stochastic uncertainties, including
errors in decision, individual trials, $etc$. $T=0$ denotes the
complete rationality, where the individual always adopts the best
strategy determinately. While $T>0$, it introduces some dynamical
randomness that there is a small probability to select the worse
one. $T\rightarrow\infty$ denotes the complete randomness of the
decision. This choice of $W$ takes into account the fact of bounded
rationality of individuals in sociology and also reflects natural
selection based on the relative fitness in terms of evolutionism.
Szab\'{o} \emph{et al.} studied the effect of dynamical randomness
$T$ on the stationary concentration of cooperators in Ref.
\cite{Szabo6}.

In a recent paper, Perc studied the evolutionary PDG by introducing
the random disorder in the payoff matrix \cite{noise1}. The reported
results indicated a resonance-like behavior that the frequency of
the cooperators reaches its maximum at an intermediate disorder.
Using a different approach, Traulsen \emph{et al.} also found that
the additive noise on the classical replicator dynamics can enhance
the average payoff of the system in a resonance-like manner
\cite{noise2}. Vainstein and Arenzon also reported that the disorder
in the underlying site diluted lattices can enhance the fraction of
cooperators \cite{disen}.

It is well known that intrinsically noisy and disordered processes
can generate surprising phenomena, such as stochastic resonance
\cite{SR}, coherence resonance \cite{CR}, ordering spatiotemporal
chaos by disorder \cite{disorder1}, disorder-enhanced
synchronization \cite{disorder2}, ordering chaos by randomness
\cite{randomness} \emph{etc}. In evolutionary games, the enhancement
of the frequency of cooperators at intermediate noise intensities
resembles the response of nonlinear systems to purely noisy
excitations.

Presently, much interest has given to evolutionary games on
complex graphs or in structured population \cite{Nowak, Hauert1,
Hauert2, netgames} by considering the fact that who-meets-whom is
determined by spatial relationships or underlying networks.
Complex networks provides a natural framework to describe the
population structure. It has been shown that, in many real-life
cases, relationships among networked individuals are neither
completely random nor completely regular, but somewhere in between
\cite{WS,BA}. In other words, real networks have some degree of
topological randomness. It is well accepted that the topology of a
network often plays crucial roles in determining the dynamics
\cite{dynamics}. Therefore, it is natural to ask whether this new
type of randomness will play some constructive roles for the
dynamics of the evolutionary games, i.e. is of benefit to the
cooperation, such as stochastic uncertainties $T$ \cite{Szabo6},
disordered payoff matrix \cite{noise1}, noise of replicator
dynamics \cite{noise2}, or disordered environments\cite{disen}.

\begin{figure}
\scalebox{0.8}[0.80]{\includegraphics{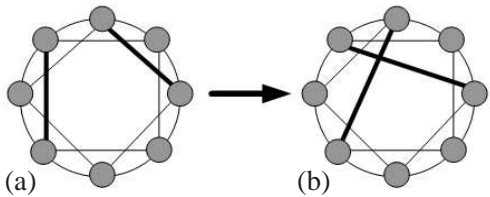}}
\caption{\label{fig:epsart} (a) Illustration of a regular ring
graph with connectivity $z=4$. Two edges are chosen and marked by
thick lines. (b) Swap the ends of the two chosen edges. The
swapped edges are marked by thick lines.}
\end{figure}

In this paper, we study the effects of both the topological
randomness in individual relationships and the dynamical
randomness in decision makings on the evolution of cooperation. We
found that there exists an optimal amount of randomness, inducing
the highest level of cooperation. The mechanism of randomness
promoting cooperations resembles a resonance-like fashion, wherein
the randomness-induced prevalence of the `good' strategy, i.e.,
cooperations, evokes the positive effect of the topological and
dynamical randomness on the system.
\begin{figure}
\scalebox{0.50}[0.50]{\includegraphics{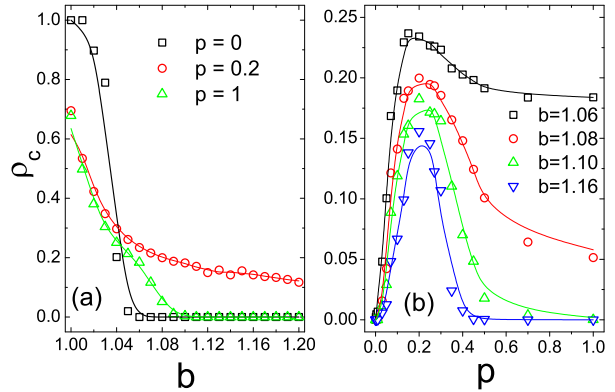}}
\caption{\label{fig:epsart} (Color online). (a) The frequencies of
cooperators $\rho_c$ vs the temptation to defect $b$ for $p=0,
0.2, 1$, respectively, with $T=0.08$. (b) $\rho_c$ as a function
of the topological randomness $p$ with various values of the
temptation to defect $b$ for $T=0.08$. The lines are used to guide
eyes.}
\end{figure}

To explore the topological randomness, we consider a homogeneous
small-world network (HSWN) \cite{HoSW}. Starting from a undirected
regular graph with fixed connectivity $z$ and size $N$, two-step
circular procedure is introduced: (i) choose two different edges
randomly, which have not been used yet in step (ii) and (ii) swap
the ends of the two edges. Here, duplicate connections and
disconnected graphs are avoided. The annealed randomness is
characterized by the parameter $p$, which denotes the fraction of
swapped edges in the network. (An illustration of the swap process
is shown in Fig. 1.) In contrast to the Watts-Strogatz (WS) model
\cite{WS}, this network has small-world effect together with
keeping the degree of each individual unchanged, so that the pure
topological randomness can be investigated by avoiding any
associated heterogeneity of degree distribution \cite{HoSW,note1}.

In all cases below, simulations start from a population of
$N=1000$ individuals located on the vertices of a regular ring
graph of $z=6$ with periodic boundary conditions. Initially, an
equal percentage of strategies (cooperators or defectors) was
randomly distributed among the population. Equilibrium frequencies
of cooperators ($\rho_c$) were obtained by averaging over $5000$
generations after a transient time of $10000$ generations. Each
data is obtained by averaging over $10$ different network
realizations with $10$ runs for each realization. Here, we adopted
a synchronous updating rule.

Figure 2 (a) shows the frequencies of cooperators $\rho_c$ on the
HSWN as a function of $b$ for different values of the topological
randomness $P$ with $T=0.08$. In the equilibrium state, $\rho_c$
is independent of the initial state and decreases monotonically as
$b$ increases. One can find that when $b<1.04$, cooperators
dominate defectors significantly on the regular ring graph ($p=0$)
and the more randomness of the network, the worse the cooperation
is. While for $b>1.04$, the cooperator is nearly extinct in the
case of $p=0$ and $p=1$, which correspond to the complete regular
network and the complete random network, respectively. However,
the cooperator can survive around $p=0.2$, i.e., intermediate
topological randomness.

The dependence of $\rho_c$ on the topological randomness $p$ is
presented in Fig. 2 (b). It illustrates that there is a clear
maximum $\rho_c$ around $p=0.2$, where cooperation can be
revitalized and maintained for substantially large values of $b$.
This phenomenon reveals that there exists somewhat resonant
behaviors reflected by the optimal cooperation level at
intermediate topological randomness, similar to the effects of
noise and disorder in nonlinear systems. However, it is worth to
mention that the dynamics leading to these equilibriums is the
same, and the resonant dependence of cooperation on $p$ results
from the changes of the equilibrium states. Moreover, in Fig.2(b),
one can find that as $b$ increases, the positive effect of
topological randomness on cooperation is restricted by the favored
defection action, which is demonstrated by the reduction of the
maximum value of $\rho_c$.

\begin{figure}
\scalebox{0.68}[0.68]{\includegraphics{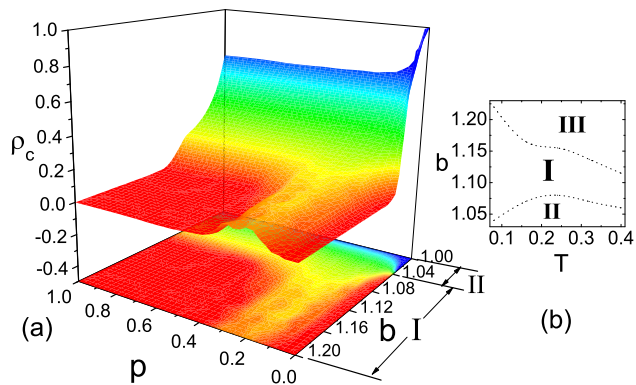}}
\caption{\label{fig:epsart} (Color online). (a) The frequencies of
cooperators $\rho_c$ vs the parameter space $(b, p)$ for $T=0.08$.
This figure illustrates two regions for $b$: I is the resonant
region where there is an optimal amount of topological randomness
$p$ enhancing $\rho_c$; II is the harmful region where the
topological randomness $p$ decreases the level of the cooperation
$\rho_c$. In fact, there exists a region III beyond the shown
range of $b$ in this figure, where the cooperators vanish and
there is no $p$ can persist or enhance the cooperation. We call it
the absorbing region. (b) The phase diagram of the three regions
of $\rho_c$ in the parameter space $(b, T)$. It illustrates that
the resonant region decreases as the dynamical randomness $T$
increases.}
\end{figure}

In the case of regular ring graph, the local spatial relationship
constrains the spreading of cooperators. However, when $p\neq 0$,
the shortcut generated by the edge-swapping reduces the average
distance of the relationship network and promotes the strategy
spreading efficiency, which induces the survival and enhancement
of cooperation. At this point, the individuals in the system can
keep clustering locally, meanwhile they can communicate each other
more effectively due to the random shortcuts. As $p\rightarrow 1$,
where the edges are exchanged sufficiently, the spatial
relationships of the individuals are completely random and the
system satisfies the mean-field approximation. Based on the
classical mean-field theory, the average payoff for the $C$ and
$D$ strategies are $M_C= z \rho_c$ and $M_D= z \rho_c b$, where
$M_D > M_C$ always since $b > 1$. According to the dynamical rule
(\ref{eq:tpo}), we can write down the following equation for the
motion of the frequency of cooperators:
\begin{eqnarray}
{\partial \rho_c \over \partial t}
&=& \rho_c ( 1-\rho_c ) [W_{D \leftarrow C} - W_{C \leftarrow D}]   \nonumber \\
&=& - \rho_c ( 1-\rho_c ) \tanh \left( {M_D - M_C \over 2T } \right)
\label{eq:em2} \;.
\end{eqnarray}
It indicates that $\rho_c$ tends to zero for arbitrary value of
$T$ as $M_D > M_C$. On the other hand, in the absence of
topological randomness $(p=0)$, the regular relationship graph can
be considered as a one-dimensional system in which cooperators
also die out \cite{oneD}. While in the optimal region of $p$, the
underlying network has the ``small-world" property: the short
average distance promotes the spreading of cooperators; the common
cluster structure induces the clustering of cooperators, leading
to the surviving and enhancement of cooperation \cite{Hauert2}.
Thus, the optimal topological randomness $p$ emerges.

To quantify the ability of topological randomness $p$ to
facilitate and maintain cooperation for various $b$ more
precisely, we study $\rho_c$ depending on $b$ and $p$ together, as
shown in Fig.3(a). One can find that when $b<1.04$, $\rho_c$ is a
monotonically decreasing function of $p$. We call it the harmful
region (denoted by $II$ in Fig.3(a)) because the topological
randomness $p$ always decreases $\rho_c$. While for $b>1.04$ (the
region is denoted by $I$ in Fig.3(a)), there exists an optimal
level of $p$ around $0.2$, resulting in the maximum value of
$\rho_c$. The positive effect of the appropriate topological
randomness $p$ on the dynamics indicates the existence of an
interesting resonance-like manner in the evolutionary game. Thus,
we call $I$ the resonant region. In fact, there exists a region
$III$ beyond the shown range of $b$ in Fig.3(a), where the
cooperators vanish and there is no $p$ can persist or enhance the
cooperation. We call it the absorbing region.

Besides the topological randomness $p$, we have studied the effect
of the dynamical randomness $T$. Figure 3(b) illustrates the phase
diagram of the three regions of $\rho_c$ in the parameter space
$(b, T)$. It shows clearly that as the dynamical randomness $T$
increases, the resonant region reduces, \emph{i.e.} the area of
the range of $b$ where the optimal $p$ can promote the cooperation
decreases, indicating the constructive effect of the optimal
topological randomness is restricted by the higher dynamical
randomness.

\begin{figure}
\scalebox{0.75}[0.75]{\includegraphics{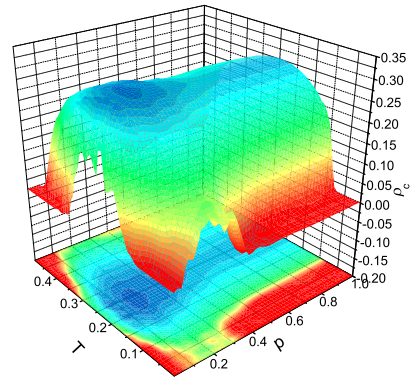}}
\caption{\label{fig:epsart} (Color online) The frequencies of
cooperators $\rho_c$ vs the parameter space $(T, p)$ for fixed
$b=1.08$.}
\end{figure}

To investigate the combined effect of both the topological
randomness and the dynamical randomness on the evolutionary
dynamics, we fix $b=1.08$, and calculate $\rho_c$ in dependence on
various $p$ and $T$, as shown in Fig.4. It is found that there
exists a clear ``optimal island" in the parameter space $(T, p)$
where $\rho_c$ reaches the highest value, indicating that the
cooperation can be promoted by both the appropriate topological
and the dynamical randomness. In other words, the resonance
induced by the dynamical randomness can be enhanced by the
topological randomness, just as the noise-induced temporal and
spatiotemporal order can be greatly enhanced by an appropriately
pronounced small-world connectivity of coupled elements
\cite{HuPerc}.

In addition, we study the PDG on the WS network \cite{WS} to give
a comparison with the cases on HSWNs \cite{note2}. In Fig.5, we
calculate the dependence of $\rho_c$ on the rewiring probability
$p_r$ of the WS model. Contrary to the results in Fig.2(b), there
is not any optimal amount of the topological randomness. Instead,
$\rho_c$ rapidly approaches a plateau at $p_r=0.2$, which is the
optimal value in the case of HSWNs. Since the only difference
between the HSWN and the WS model is that the variance of the
degrees in the latter is nonzero \cite{HoSW}, we conclude that
when $p_r$ is over 0.2, $\rho_c$ can be enhanced considerably as a
result of the redundant heterogeneity on relationships among
individuals. Moreover, due to the additive heterogeneity, the
plateau in Fig.5 is clearly higher than the corresponding maximal
values in Fig.2(b). Thus, the plateau of $\rho_c$ is the combined
effect of the topological randomness and the heterogenous spatial
relationships.

\begin{figure}
\scalebox{0.55}[0.55]{\includegraphics{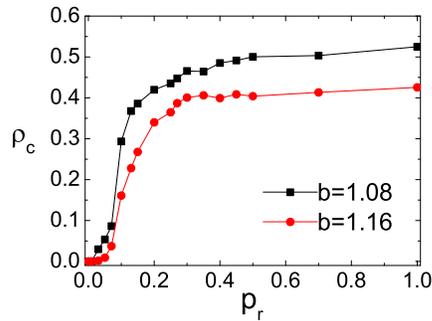}}
\caption{\label{fig:epsart} (Color online) The frequencies of
cooperators $\rho_c$ as a function of the rewired probability
$p_r$ with $b=1.08, 1.16$ for $T=0.08$. The lines are used to
guide eyes.}
\end{figure}

To summarize, we have studied the effects of both the topological
randomness and the dynamical randomness on the evolutionary
Prisoner's Dilemma game and found that there exists an optimal
amount of randomness, leading to the highest level of cooperation.
The mechanism of randomness promoting cooperation resembles an
interesting resonance-like phenomenon, wherein the
randomness-induced prevalence of the cooperation evokes the
positive role of the topological and dynamical randomness in the
system. Moreover, we find that the heterogeneity in the underlying
relationship net also enhances the cooperation. Although our work
is exclusively based on the evolutionary Prisoner's Dilemma game,
the ``resonant" behavior may play a significant role in other
styles of evolutionary dynamics.

This work is funded by NNFC No.10635040. We thank M. Perc for
valuable discussion.


\begin{thebibliography}{10}
\bibitem{cooperate1} A.M. Colman, {\sl Game Theory and its Applications in the Social and Biological
Sciences} (Oxford, 1995); J. Hofbauer and K. Sigmund,  {\sl
Evolutionary Games and Population Dynamics} (Cambridge, U.K.,
1998).

\bibitem{kin} W.D. Hamiltion, J. Theor. Biol. \textbf{7}, 17 (1964).
\bibitem{Axelrod} R. Axelrod and W.D. Hamilton, Science
\textbf{211}, 1390 (1981); R. Axelrod, {\sl The Evolution of
Cooperation} (Basic books, New York, 1984).
\bibitem{Nowak1} M.A. Nowak and K. Sigmund, Nature \textbf{393}, 573 (1998).
\bibitem{Hauert1} C. Hauert et al., Science \textbf{296}, 1129 (2002).
\bibitem{reputation} E. Fehr and U. Fischbacher, Nature \textbf{425}, 785 (2003).
\bibitem{Nowak} M.A. Nowak and R.M. May, Nature \textbf{359}, 826 (1992).

\bibitem{Szabo5} G. Szab\'{o} and C. T\"{o}ke, Phys. Rev. E
\textbf{58}, 69 (1998).

\bibitem{May} M.A. Nowak and R.M. May, Int. J. Bifurcation Chaos \textbf{3},
35 (1993).

\bibitem{Szabo6} G. Szab\'{o}, J. Vukov, and A. Szolnoki, Phys. Rev. E
\textbf{72}, 047107 (2005); J. Vukov, G. Szab\'{o}, and A. Szolnoki,
Phys. Rev. E \textbf{73}, 067103 (2006).

\bibitem{noise1} M. Perc, New J. Phys. \textbf{8}, 22 (2006).

\bibitem{noise2} A. Traulsen, T. R\"{o}hl, and H. G. Schuster, Phys.
Rev. Lett. \textbf{93}, 028701 (2004).

\bibitem{disen} M.H. Vanistein and J.J. Arenzon, Phys. Rev. E
\textbf{64}, 051905, (2001).

\bibitem{SR} L. Gammaitoni, P. Hanggi, P. Jung, and F. Marchesoni,
Rev. Mod. Phys. \textbf{70}, 223 (1998).

\bibitem{CR} A.S. Pikovsky and J. Kurths, Phys. Rev. Lett. \textbf{78}, 775 (1997).

\bibitem{disorder1} Y. Braiman, J.F. Linder, and W. L. Ditto,
Nature \textbf{378}, 465 (1995).

\bibitem{disorder2} K. Wiesenfeld, P. Colet, and S. H. Strogatz,
Phys. Rev. Lett. \textbf{76}, 404 (1996).

\bibitem{randomness} F. Qi, Z.-H. Hou, and H.-W. Xin, Phys. Rev.
Lett. \textbf{91}, 064102 (2003).

\bibitem{Hauert2} C. Hauert, and M. Doebeli, Nature \textbf{428}, 643 (2004).



\bibitem{netgames} E. Lieberman, C. Hauert, and M.A. Nowak, Nature \textbf{433}, 312
(2005); H. Ohtsuki et al., Nature \textbf{441}, 502 (2006); F.C.
Santos and J.M. Pacheco, Phys. Rev. Lett. \textbf{95}, 098104
(2005); F.C. Santos, J.M. Pacheco, and T. Lenaerts, Proc. Natl.
Acad. Sci. USA \textbf{103}, 3490 (2006); G. Szab\'{o} and C.
Hauert, Phys. Rev. Lett. \textbf{89}, 118101 (2002); W.-X. Wang et
al., Phys. Rev. E \textbf{74} 056113 (2006).



\bibitem{WS} D.J. Watts and S.H. Strogatz, Nature (London)
\textbf{393}, 440 (1998).

\bibitem{BA} A.-L. Barab\'asi and R. Albert, Science
\textbf{286}, 509 (1999).

\bibitem{dynamics} S. Boccaletti, V. Latora, Y. Moreno, M. Chavez, and D.-U.
Hwang, Physics Reports \textbf{424}, 175 (2006).

\bibitem{HoSW} F.C. Santos, J.F. Rodrigues, and J.M. Pacheco, Phys. Rev. E \textbf{72}, 056128 (2005).

\bibitem{note1} The HSWN possesses small-world effect together with
identical degree of nodes. In opposite, the heterogeneity
indicates the difference of nodes' degrees.

\bibitem{oneD} G. Szab\'{o}, T. Antal, P. Szab\'{o}, and M. Droz, Phys. Rev. E \textbf{62}, 1095
(2000).

\bibitem{HuPerc} Z. Gao, B. Hu and G. Hu, Phys. Rev. E
\textbf{65}, 016209, (2001); M. Perc, New J. Phys. \textbf{7},
252, (2005)

\bibitem{note2} Details of WS model can be seen in Ref. \cite{WS}.


\end{thebibliography}
\end{document}